# REACT: REActive resilience for critical infrastructures using graph-Coloring Techniques

Ivan Marsa-Maestre[a], Jose Manuel Gimenez-Guzman[a], David Orden[b], Enrique de la Hoz[a], Mark Klein[c]

[a]*Computer Engineering Department, University of Alcalá, Spain*
[b]*Department of Physics and Mathematics, University of Alcalá, Spain*
[c]*Center for Collective Intelligence, Massachusetts Institute of Technology, Cambridge (MA), US*

ABSTRACT: Nowadays society is more and more dependent on critical infrastructures. Critical network infrastructures (CNI) are communication networks whose disruption can create a severe impact. In this paper we propose REACT, a distributed framework for reactive network resilience, which allows networks to reconfigure themselves in the event of a security incidents so that the risk of further damage is mitigated. Our framework takes advantage of a risk model based on multilayer networks, as well as a graph-coloring problem conversion, to identify new, more resilient configurations for networks in the event of an attack. We propose two different solution approaches, and evaluate them from two different perspectives, with a number of centralized optimization techniques. Experiments show that our approaches outperform the reference approaches in terms of risk mitigation and performance.

## 1. Introduction

Cyber-attacks over critical infrastructures have proliferated at an alarming rate in the last decade [1]. This poses a significant threat, due to the high level of dependency that our society has on what are known as critical infrastructures. If any of these infrastructures is compromised, it can have a serious impact in our life and disrupt the normal society order as a whole. The protection of critical infrastructures against cyber-attacks is thus a top priority for governments and critical service operators.

Critical infrastructures are characterized by a high level of interconnection. Many of these (physical, logical, or virtual) dependencies are not revealed until a crisis arises. This high level of interdependency may lead to cascading failures. Even small disruptions can be enough to unleash dramatic consequences in high-complexity systems.

The entry point for most of these cyber-attacks are what we call critical network infrastructures (CNI), which are those communication network infrastructures whose disruption, intentional or accidental, has a high impact because of either its massive, even global, deployment (e.g., Internet, cell provider networks) or its supporting role for other critical infrastructures (e.g., the internal communications network for the control system in a nuclear facility, or the communication network for the electrical grid).When addressing the challenge of increasing robustness in CNI, the concept of network resilience emerges. We can define network resilience [2, 3] as the ability of a network to defend itself and keep an acceptable service level in the presence of such challenges as malicious attacks, hardware failures, human mistakes (hardware or software misconfigurations) and large-scale natural disasters that may threaten its normal operation.

In this work, we address *reactive network resilience*, which extends the concept of network resilience to cover the ability of a network to reconfigure itself in the event of a security incident so that the risk of further damage is mitigated. This is radically different from the usual notions of resilience, where the network is designed so that the attack impact is minimized a priori. For instance, [4] designs cloud-based critical infrastructures so that there are redundancies which minimize the impact of a service disruption. In [5, 6] the network is designed so that diversity is maximized, to prevent attacks which affect similar configurations to propagate throughout the network. In contrast, we take advantage of the possibilities of new (semi-)virtualized network infrastructures to perform the network configuration in a reactive manner once the attack has been detected. In this way, the resulting configuration is specific for the location of the attack, hence allowing for a greater risk reduction.

In particular, we propose REACT, a distributed framework for reactive network reconfiguration in (semi-)virtualized CNIs over a zero-day attack threat model. This threat model assumes that the specific vulnerabilities used by the attack are unknown to the network operators, and therefore can only be detected by using an anomaly-based IDS (Intrusion Detection System). REACT uses the alert reports from such an IDS, as well as an *a priori* risk analysis model of the network, to propose a new network configuration (that is, an alternative redeployment of the different components in the network) that will minimize the impact of the incident. It is important to note that, although malicious attacks are not the only challenge to cope with in CNIs, these types of attacks are generally the most harmful ones as they are focused on blocking network services. Network resilience challenges are also different with malicious attacks, human errors or natural disasters [7]: the latter two have a more random and distributed nature, while the former are usually focused on nodes that play an essential role in the network. A more in-depth discussion of related literature, with an emphasis on zero-day resiliency, can be found in Section 2.

Our goal is, therefore, to build a distributed, reactive risk-mitigation framework for (semi-)virtualized critical network infrastructures. The paper contributes to this goal in the following ways:

- We propose a risk model, based on multi-layer networks, which takes into account conventional risk analysis metrics and novel zero-day resilience metrics (Section 3.1).
- We propose two different perspectives to make risk mitigation, using this risk model, tractable in real time (Section 3.2).
- We propose a distributed system for reactive redeployment management, with a strategy based on a solution the well-known graph coloring problem (Section 3.3).
- We adapt several state of the art techniques to identify how to redeploy networks upon a security incident alert (Section 3.3.2), and propose two novel approaches which benefit from the graph structure of the problem (Section 3.3.3).

To validate our approach and evaluate the impact of our contributions, we conducted a set of experiments over a range of randomly-generated scenarios. This is described in Section 4. Our results show that our approaches outperform the reference techniques in terms of both risk mitigation and performance. In Section 4.3, we report our evaluations of the effect of some network properties on risk reduction. In section 5, we summarize our contributions and identify lines for future research.

The rest of the paper is structured as follows. The next section includes the related work while Section 3 describes the proposed framework for network resilience in critical infrastructures, defining the model and the techniques used for its optimization. The performance evaluation and results of the proposal are shown in Section 4. Finally, a summary of the paper and some concluding remarks are given in Section 5.

## 2. Related Work

### 2.1. Network resiliency under zero-day attacks

Recently, there has been a growing interest in maximizing the resilience of critical infrastructures [8], which increasingly rely on networked computer systems. Network vulnerability is a widely acknowledged problem with a history of global scale incidents, both accidental [9] and as a result of malicious activities [10]. In this work we focus on malicious activities and, more specifically, on zero-day attacks. While for already-known attacks the solution will be probably found in the use of firewalls, antivirus or software patching and upgrading, the mitigation of zero-day attacks is a harder task due to the lack of information. In the literature, the work related to zero-day attacks comprises two different fields. The first one is the detection of zero-day attacks [11, 12]. The second one, more related to this work, is devoted to the development of metrics for evaluating the resilience of a network against zero-day attacks. The proposal of metrics is of vital importance to measure the network security, as "you cannot improve what you cannot measure" [13]. Network security metrics usually assign scores to vulnerabilities according to their probability and impact. A review and performance comparison of such metrics can be found in [14-17]. However, this type of metrics is not valid for zero-day attacks, due to the lack of knowledge about them. For that reason, there are some works ([18] and [13]) that propose metrics specifically suited for zero-day attacks. In [18], authors consider network diversity as a metric for evaluating resilience, while in [13] authors propose a metric based on counting how many distinct zero-day vulnerabilities are required to compromise a network asset.

Another interesting field related to our work, but not specially devoted to zero-day attacks, are the pioneering contributions [19] and [20], where authors analyze the robustness of complex networks to localized attacks, to determine how much damage a network can sustain before it collapses. The differences between these efforts and our own are clear, as these papers are more theoretical, and we consider more realistic computer network models. Our network models are more realistic, we believe, because they are based on multi-layer graphs. It has become increasingly clear that we need to go beyond monolayer graph modelling and explore more realistic and complex models. In [21], for example, resilience in cloud computing is studied using multilayer graphs. Multiplex or multi-relational networks connect nodes using links that can express different kinds of relationships [22]. Multilevel networks and meta-networks enable hierarchical structures and nodes and links of different types [23]. Recently, [24] has presented a unified modelling framework for multi-layer networks that includes these concepts in a unified manner that takes advantage of the different mathematical tools available in the state of the art.

The last scientific field to which our work is related to, and also we intend to contribute, involves the use of optimization techniques based on negotiation to maintain network resilience in a distributed and self-organized way. These optimization techniques determine the optimal design of critical networks, in terms of the component configuration and network topology, to maximize resilience against zero-day attacks. Traditionally, networks are deployed to fulfil business requirements, while network security is left as an afterthought. More recent work suggests, however, that network configuration and topology should play a paramount role in network design, even to the extent of creating a dynamic environment to hinder the purpose of attackers [25]. This approach of reconfiguring the network topology has been successfully deployed in different scenarios like Internet of Things (IoT) [26].

The distributed and adaptive nature of critical network infrastructures, along with the need to reach a consensus between conflicting individual goals to achieve a social goal, suggests the use of negotiation techniques. However, most negotiation research has focused on problems with one issue (typically price) or a few independent issues [27] and are demonstrably sub-optimal for negotiations with multiple interdependent issues [28]. While some previous efforts [29, 30] have attempted to address this challenge, they face serious limitations in terms of outcome optimality, strategic stability and scalability. These three performance indicators are key enablers for the success of optimization systems in large real-world infrastructures, due to their continuous increase in network size, structural complexity and dynamicity [31]. One of the contributions of this paper is to provide a realistic setting where the possibilities for application of these techniques can be evaluated.

*2.2. (Semi-)virtualized networking*

The approach described in this paper takes into account the ongoing transformation of datacenters and datacenter networks, driven by virtualization, automation, and containers. Virtualization has transformed the way in which we design and manage servers and datacenters. This has led to increasing efficiencies and a trend towards automation. Virtualization along with configuration management tools, such as Puppet, Chef or Ansible [32], has led to a new approach known as programmable infrastructure or Infrastructure as Code (IaC) [33]. Infrastructure as Code is an approach to infrastructure automation, based on practices from software development, that emphasizes consistent, repeatable routines for provisioning and changing systems and their configuration. IaC can be defined as "the process of managing and provisioning computing infrastructure (processes, bare-metal servers, virtual servers, etc.) and their configuration through machine-processable definition files, rather than physical hardware configuration or the use of interactive configuration tools" [34]. This way, IT infrastructure supports and enables change, and users are able to define, provision, and manage the resources they need. Changes to the running systems become routine and improvements can be made continuously: there is a document describing the deployment and introducing changes to the deployment is just a matter of modifying that document and applying the modification by using the proper provisioning and configuration management tools. It is important to notice that this is not only limited to server management but to network creation, configuration and management, thanks to approaches and technologies such as Network Function Virtualization (NFV) and Software Defined Networking (SDN) [35].

Finally, for working with large distributed infrastructures, the use of container technologies, such as Docker [36], is also increasing. Containers are isolated places where an application can run without affecting the rest of the system, and without the system affecting the application [37]. By using containers, it is possible to isolate pieces of a system into separate containers. For instance, you can have a container for Apache, a container for MySQL, and one for MongoDB. The underlying idea is being able to create your distributed system by combining containers that provide the functionality that you need. This allows us to isolate individual elements of the application into independent units that can be managed in a flexible manner. Containers can be created and distributed by means of public or private repositories [38], just like code is distributed in places such as *github* or *bitbucket*. Container technology represents the technology foundation for the component layer that we propose in this work.

*2.3. Distributed approximations to security and critical infrastructures*

As inherently connected and interdependent complex systems, critical infrastructures are not alien to distributed modelling [39]. In fact, a number of successful approaches to security in these environments have emerged in the last years. Of particular interest are the approximations based on the idea of "security games" [40]. These efforts use computational game theory to build decision support systems for efficient security resource allocation in surveillance scenarios (e.g. airport security), by modelling the allocation process as a Bayesian Stackelberg game [40]. In domains more related to communication networks, we can find game-theoretic works regarding intrusion detection, especially for distributed systems [41]. There are some works which study virtualized scenarios, such as cloud computing [42] or software-defined networks [43]. However, both works address a very different threat (distributed denial-of-service attacks, DDoS), and they focus more on detection than on mitigation. In [42] authors try to maximize IDS detection probability, and they model the problem as a security game against the attacker, which attempts to minimize this probability. In our work we take as a starting point the detection of the compromise, and we study reactive network reconfiguration in response to that incident so that the risk of further damage is mitigated. [43] is a bit closer to our approach in the sense that it considers attack mitigation strategies, but the attack mitigation strategies are specific to DDoS attacks and SDN controller rules (e.g. forward, drop, modify), not considering the re-instantiation of components or the dynamic topological changes we deal with in this work.

Although there are works which address resiliency in cloud-based critical infrastructures [4], they do so more in the sense of providing redundancy against service disruption (specially in cyber-physical control systems). Our work, however, focuses on preventing threat propagation throughout the network, where redundant pathways may indeed jeopardize resilience rather than increasing it. This is related to the concept of network diversity [13,18], which we discussed in section 2.1.

To the best of our knowledge, our work is the first attempt to use network diversity to provide reactive resilience in the event of a cyber-attack. We did a preliminary exploration of this line of work in the conference paper [44]. After the publication of that work, other authors have recently taken a similar approach [45], although with some key differences. First of all, they study the "resilience-by-design" problem, in the sense that they try to optimize the a priori network diversity so that the overall vulnerability of the network is mitigated. In addition, they focus on diversity of security mechanisms (SM) in the protection of Smart Grids, while our model is generic and takes into account all elements of the network infrastructure. Finally, the modelling they use is a flat graph rather than a multi-layer network and does not allow for topological or deployment changes as part of the optimization process.

In this paper we deepen the work we initiated in [44], introducing significant differences in the model, approach and evaluation. First, the proposed multilayer model in this paper is more accurate, including a new layer called the instance layer as well as the possibility for layer multiplicity, which greatly increases the expressiveness of the model. Second, we introduce a more refined model for risks, including inherent risks. Third, in [44] we proposed the configuration-spread projection, and now we propose the network-spread projection, which allows improved risk reduction. Fourth, we propose a new technique to tackle this problem and compare it with the proposal made in [44], showing that the new proposal is much more efficient. Finally, the performance evaluation conducted in this paper is more comprehensive, including an analysis of the effect that network metrics have on the achievable risk reduction, and an experimental comparison which the aforementioned work [45], which shows significant advantages for our approaches in terms of risk reduction.

## 3. A Distributed Framework For Reactive Resilience In Critical Infrastructures

In this section, we describe our proposal for reactive resilience in (semi)virtualized critical network infrastructures. The framework starts from the following environment assumptions:

- There is a weighted risk assessment of the system prior to the operation of our framework. This risk assessment comprises an inventory of system assets, along with their value and interdependencies.
- Assets rely on components to function. These components can be instantiated according to different configurations.
- The main threat considered are zero-day attacks, which are normally configuration-specific, that is, they affect just a single configuration or a set of similar ones.
- Components are virtualized and can be deployed on different physical hosts along the physical network infrastructure, using a container paradigm as discussed above.
- There is a control plane where the deployment decisions are made. This control plane is in an overlay, isolated network, and thus it is assumed not to be under the discussed threats.
- There is an intrusion detection system (IDS) able to identify compromised components. We assume it to be an anomaly-detection based IDS, since zero-day attacks are by definition unknown, and therefore no additional information about the incident will be provided.

In the following, we describe our proposal, which is based on three key elements: a multilayer network model, a reactive resilience risk model which induces two alternative perspectives of the problem, and an optimization process, inspired by graph coloring, for reactive redeployment. We then describe the candidate techniques we have evaluated to guide this optimization process. First, we describe several centralized optimization techniques from the literature that we use as references. Then, we propose two novel techniques which take advantage of the graph structure of the problem: TSC-DSATUR and Gossip-Partitioned Belief-Propagation (GPBP).

## 3.1. REACT Multilayer network model

As described above, REACT takes risk assessment and alert reports and derives a new network configuration intended to maximize resilience. To achieve this goal, REACT uses a multilayer network model to capture the situational awareness input for the critical infrastructure. An outline of the model is shown in Fig. 1.

The model defines four types of layers, which may appear with variable multiplicity. The *asset layer*, on top, is extracted from risk analysis, and captures the relative importance of the assets and their interdependencies. The bottom layer is the *infrastructure layer*, which represents the actual infrastructure elements (hosts, links and network appliances) upon which the assets are deployed within the network. Between these two end layers, there are a number of component-instance layer pairs, which represent the components (e.g. databases, backend or frontend elements) upon which assets depend on to provide their functionality, and the actual instantiation of these components with particular configurations and degrees of redundancy. In the following, we describe in more detail the different layers and the information they provide to the model.

### 3.1.1. Asset Layer

As stated above, the asset layer is extracted from risk analysis, and is intended to capture the value of their different assets and the dependency relationships among them. This is represented as a weighted directed graph, where nodes represent assets and incoming edges to a node represent assets the node depends on. For instance, in the asset layer shown at the top of Fig. 1, asset $a_2$ depends on assets $a_1$ and $a_3$, asset $a_4$ depends on $a_3$, and assets $a_5$ and $a_6$ are mutually interdependent.

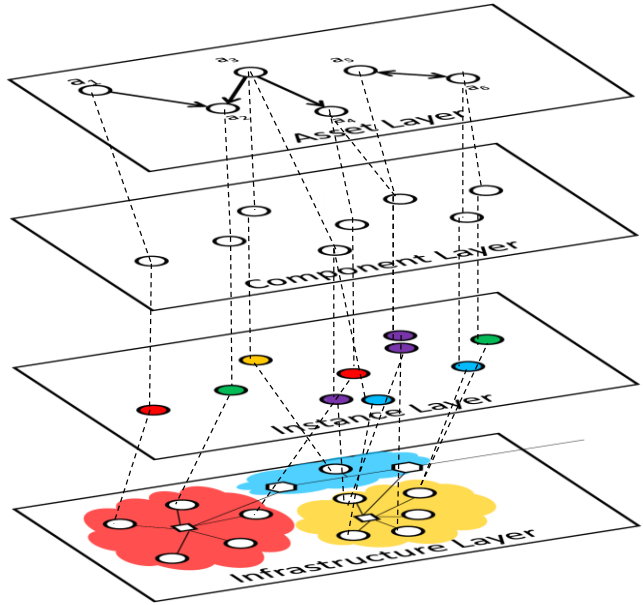

**Fig. 1.** REACT multilayer network model.

Node weights (e.g., $w_{a_1}$) are mandatory and represent the relative importance of a given asset for the organization, from the point of view of risk analysis. Edge weights (e.g., $w_{a_1 a_2}$) are optional and represent the criticality of the dependency between two assets. There may be different ways to estimate the value of these weights, but in general they will represent the relative impact on one asset of the failure of another. For instance, an edge weight $w_{a_1 a_2} = 0.5$ indicates that a failure of asset $a_1$ will reduce the value/performance of asset $a_2$ to 50%, or that if asset $a_1$ fails there is a probability of 50% of failure propagation to $a_2$.

*3.1.2. Components and instance layers*

These layers capture the components (generally software) each asset relies on to provide its functionality and value to the organization. For instance, a given control application within the critical infrastructure may require using a Web Service, or a given database. Unlike [44], the layer is doubled, with the upper layer representing the components' abstract functionality, and the lower layer representing the actual instantiation of this functionality in the system.

The reasons for this distinction are twofold. On one hand, a given component may often have multiple instances, mainly for redundancy. On the other hand, while the component layer captures the abstract functionality of a given component (e.g., a web application), the instance layer represents the actual configuration of that component (e.g., a PHP web application using a MySQL database). This double-layered approach allows us to capture a wide variety of the component instantiation policies used in critical infrastructures. For example, a component could have two instances with identical configurations or with different configurations for increased zero-day resilience. Note that, for convenience, we represent the different component configurations as colors.

Furthermore, additional component-instance layer pairs can be included in the model to provide finer granularity in the description of the component configurations. For instance, the configuration for a given web application can be provided in a single layer, with possible values being e.g., LAMP (Linux, Apache, MySQL and PHP) or WAMP (Windows, instead of Linux). Another option could be to split the configuration description into four layer pairs: operating system (Linux, Windows, MacOS…), database (MySQL, PostgreSQL…), Web Server (Apache, Nginx…) and Web Application framework (PHP, JSP…). This allows for great expressiveness when describing the system.

*3.1.3. Infrastructure layer*

This layer represents the actual physical/logical network infrastructure in the critical infrastructure, that is, the actual hosts and network elements in the infrastructure. We assume a (partially) virtualized network, so that the topology of this layer may not represent an actual physical topology. This layer could be split into virtual and physical network layers as necessary, though. For the purpose of our resilience analysis, we will assume that the network infrastructure is divided into segments (e.g., DMZs, VLANs), and we will give them colors for convenience. It is important to note that this layer also defines the entry points to the network.

*3.2. REACT reactive resilience model*

REACT will deal mainly with resilience to zero-day attacks. Zero-days are normally configuration-specific (e.g., a zero-day exploit for Mac OS X 10.10.1), and they may lead to cascade attacks (e.g., when having adjacent firewalls which are all vulnerable to the same zero-day). Therefore, we need to update the model to take into account the effect of potential vulnerabilities and attack chains, as a result of the interaction both between configurations and between topology segments. Note that in [44] only cross-segment and cross-configuration vulnerabilities were taken into account, while we now also consider inherent vulnerabilities.

To account for the impact, on resilience, of using specific configurations in our system, we include two parameters:

- **Inherent zero-day vulnerability** ($v_i$): for a given configuration $\gamma_i$, this vulnerability value $v_i$ accounts for the relative probability of a zero-day being discovered for that particular configuration. This probability can be estimated from previous advisories or from other risk analysis tools and reports. For instance, recent security reports may indicate that the existence of a zero-day exploit against a LAMP configuration is very low, and therefore we can assign such configuration an inherent zero-day vulnerability value of 0.05.
- **Cross-configuration zero-day vulnerability** ($v_{ij}$)**:** this vulnerability value $v_{ij}$ accounts for the probability that a zero-day exploit over configuration $\gamma_i$ would allow also to compromise configuration $\gamma_j$. Usually, this is directly related to configuration diversity [46]. For instance, given that LAMP and WAMP configurations are extremely similar, they could be assigned a cross-configuration zero-day vulnerability value of 0.8.

*3.2.1. Topology-related vulnerabilities*

These vulnerabilities are related to the ease by which an attacker can reach a specific segment of the network infrastructure, or jump from one network segment to another. This, in turn, is directly related to the routing policies as well as the access control rules configured in firewalls. From the analysis of these rules, we can derive, as before, two different vulnerability values:
- **Inherent segment vulnerability** ($\mu_i$)**:** for a given topology segment $s_i$, this vulnerability value $\mu_i$ represents how easy would be for an attacker to send packets to this segment of the network. For instance, the inherent vulnerability value for the outmost DMZ of the network could be 1 (could be lower, but since it is used only in relative terms, that would be indifferent), while the vulnerability value for an adjacent network segment protected by a firewall could be 0.8.
- **Cross-segment vulnerability** ($\mu_{ij}$)**:** this vulnerability value $\mu_{ij}$ represents how easy would be for an attacker to send packets from segment $s_i$ to segment $s_j$, provided that a machine in segment $s_i$ has been compromised. Again, this largely depends on firewall access control rules, though it can be generally assumed that adjacent segments will have higher cross-segment vulnerability values than distant ones.

*3.2.2. Potential attack paths from a given security incident: the network-spread projection*

REACT is intended to act in response to an alert from an IDS. Such alert will correspond to anomalous behavior in a given component, host or network segment. Both machine-wide or segment-wide alerts can be generalized to component-wise alerts, by assuming a worst-case scenario where all components in the affected machine or network segment may have been compromised. Therefore, without loss of generality, we can assume that REACT reacts to compromises in components. If a component has been compromised, *attack paths* can be traced from these components to other components in the system, provided that there is a way an attacker could progress from one component to another, throughout the network and by compromising successive machines. To take this into account, we can "flatten" the multilayer network model into a *network-spread projection* from a given compromised component.

Figure 2 shows an example of such a projection. In this example, component $C_1$ has been compromised, and

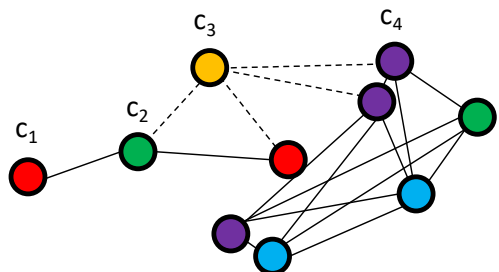

**Fig. 2.** Network-spread projection of the model.

the projection shows all components which are potentially reachable from that node. Two components $c_i$ and

$c_j$ are connected if there is a network path between them (that is, if an attacker could send packets from the network segment $s_i$ containing component $c_i$ and the network segment $s_j$ containing component $c_j$). The weight of the edge connecting these two nodes would be the corresponding cross-segment vulnerability value $\mu_{ij}$. In the figure, lower-weight edges are represented in dashed lines for convenience, and instance configurations are represented by colors.

With this projection in mind, an *attack path* from any node (for example, $c_1$) to another node (say $c_4$, for instance), is a loopless path between these two nodes (in our example, for instance, path $c_1 c_2 c_3 c_4$). Any attack path $P$ will have an associated *attack probability* $\propto_P$, computed as the product of the edge weights in the path (given by $\mu_{ij}$) and the *node contagion costs* given by the minimum of the cross-configuration vulnerability value $v_{ij}$ among the already traversed nodes in the path, to account for the fact that any already compromised (or similar enough) configuration will be easy to compromise for an attacker. That is, for a given attack path $P = (c_1, c_2, \ldots, c_N)$, the associated probability $\propto_P$ will be:

$$\propto_P = \prod_{i=1}^{N-1} \left( \mu_{i,i+1} \cdot \min_{j<i+1} v_{j,i+1} \right)$$

This allows us to estimate the worst-case overall risk over a given component $c_i$ when we know that $c_j$ has been compromised as the risk of the highest probability attack path:

$$\rho_{c_i|c_j} = \max_{\forall P_{c_j \to c_i}} \propto_P$$

Given that every component $c_i$ is associated to an asset $a_i$ (and we denote it as $c_i \to a_i$) through the links between the asset layer and the component layer, we can compute the inherent risk over a given asset after the compromise of $c_j$ as

$$\rho_{a_i|c_j} = \sum_{k|c_k \to a_i} \rho_{c_i|c_j}$$

Taking into account the dependencies between assets $w_{a_k \to a_i}$, extracted from the asset layer, we can iteratively compute the aggregated risk over all assets $a_i$, as follows:

- At iteration 0: $\overline{\rho_{a_i|c_j}}^0 = \rho_{a_i|c_j}$
- At iteration t: $\overline{\rho_{a_i|c_j}}^t = \rho_{a_i|c_j} + \sum_{k|a_k \to a_i} w_{a_k \to a_i} \overline{\rho_{a_k|c_j}}^{t-1}$

This computation is guaranteed to converge in at most $d$ iterations, where $d$ is the diameter of the graph induced by the asset layer, yielding a final value for the risk over each asset. From these risks and the asset weights (also from the asset layer), we can finally derive the overall risk over the system as a result of the compromise of component $c_j$:

$$\overline{\rho_{c_j}} = \sum_{a_i} w_{a_i} \overline{\rho_{a_i|c_j}}^d$$

Therefore, the goal of REACT is to produce an alternative configuration of the system which minimizes this risk.

*3.2.3. An alternative perspective of the problem: the configuration-spread projection*

The model above, though correct, imposes some computational overhead which limit its scalability. Ignoring the effect of the cross-configuration vulnerability values, $\rho_{c_i|c_j}$ computation can be thought as an analogy of the shortest path problem with non-negative real weights (taking $1/\mu_{ij}$ as edge weights), which is roughly solvable in $O(E + V)$, with $E$ and $V$ being, respectively, the edges and vertices of the network-spread projection. However, when we introduce the effect of the cross-configuration vulnerability values, the way $\propto_P$ is defined requires to compute $\min_{j<i+1} v_{j,i+1}$ for every attack path, which is in the worst case $O(E^2)$. Given that REACT is expected to respond to the attack events in a timely fashion, such model is not useful for this scenario.

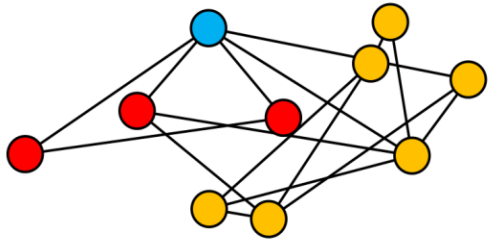

**Fig. 3.** Configuration-spread projection.

Taking this into account, we propose an alternative projection of the multilayer network model, which we call *configuration-spread projection*. In this case, components are connected if there exists a significant similarity between their configurations which involves the risk of zero-day spread, and the weight of the links is the cross-configuration vulnerability value $v_{ij}$. Figure 3 shows the corresponding projection for our example, where we have connected those nodes with similar configurations. In this projection, node color represents the network segment where a particular component instance has been deployed, according to the infrastructure layer.

With this projection in mind, risk computation is identical to the previous section, except for the computation of the *attack probability* $\propto_P$ of a given attack path. Now, this probability is computed as the product of the edge weights in the path (given by $v_{ij}$) and the *node traversal costs* given by cross-segment vulnerability value $\mu_{ij}$. This yields the following expression for $\propto_P$ for a given attack path $P = (c_1, c_2, \ldots, c_N)$:

$$\propto_P = \prod_{i=1}^{N-1} \mu_{i,i+1} \cdot v_{i,i+1}$$

This problem reformulation, though theoretically equivalent to the previous one, is much more adequate for REACT. The reasons for this are twofold. On one hand, this more compact perspective reduces the computational complexity of $\rho_{c_i|c_j}$, which in turn reduces REACT's reaction time. On the other hand, for a given graph structure of the projection, network reconfiguration after an attack reduces to *re-coloring the graph* -that is, redeploying components in an alternative topology- so that the overall risk induced by the security incident is minimized. This allows us to explore a range of successful techniques for topology optimization, as we will see in the following sections.

### 3.3. REACT framework implementation

#### 3.3.1. Spectrum coloring for network reconfiguration

From the two projections (network-spread and configuration-spread) discussed above, we can see that network reconfiguration in the context of REACT (that is, in reaction to an alert) is a nonlinear optimization process that involves finding the configuration which minimizes the overall risk $\overline{\rho_{c_j}}$ caused by the incident. General optimization of $\overline{\rho_{c_j}}$, as we will see in the evaluation, is a costly process in terms of computation time, and is therefore poorly-suited for the real-time performance requirements of the REACT system. It is necessary, therefore, to find approaches which can yield satisfactory (albeit probably sub-optimal) solutions in a reasonable time.

To accomplish this, we will adopt a simplification of the problem, whose rationale is as follows. Given that $\overline{\rho_{c_j}}$ depends on the maximum probability $\propto_P$ for all attack paths, and that the factors contributing to that probability are the cross-segment vulnerability values and the cross-configuration vulnerability values of adjacent nodes in the projections (Fig. 2 and Fig. 3), intuitively we need to minimize these vulnerability values in adjacent nodes. Roughly speaking, we need to avoid having highly similar configurations in highly reachable network segments, to make it difficult to trace successful attack paths through the network.

This aim of avoiding similar features in close items is similar to that of the mathematical graph coloring problem, see [47] and the references therein. In this well-studied problem, an abstract graph (a set of nodes together with edges defined as pairs of nodes to be connected) and a set of colors are given, with the aim of assigning a color to each node in such a way that connected nodes do not get the same color. This idea has been proved particularly useful for frequency assignment where, colors standing for frequencies and connected nodes denoting potential interferences, such a coloring ensures that interferences are avoided. However, in our problem we need a more flexible coloring model which, instead of just forbidding the same configurations in connected items of the network, aims to avoid highly similar configurations in highly reachable items. Thus, instead of the classical graph coloring problem we will consider the *Threshold Coloring Problem* (TSC) [48], where we also have an abstract graph and a set of available colors, but in addition we consider an interference matrix which assigns an interference value to each possible way of coloring a pair of connected nodes.

The goal of this TSC problem is to find a coloring which minimizes the maximum interference per node (the optimal solutions for the example problem can be seen shadowed in the figure). Our hypothesis is that, by applying the techniques we have used successfully for this problem to the problem of recoloring the configuration-spread projection (as we stated in the previous section), we will find suitable resilient alternative topologies in a reasonable time. Of course, the translation from one problem to the other is not direct, but it is fairly viable. With the perspective of the configuration-spread projection, the color set represents the different network segments $s_i$ where we can deploy the different components, and the color interference matrix represents the corresponding cross-segment vulnerability values $\mu_{ij}$. We also have to augment the model to introduce edge weights, which correspond to the cross-configuration vulnerability values $v_{ij}$ of the components represented by adjacent nodes.

Alternatively, from the network-spread projection perspective, the colors can represent the different configurations for components, the color interference matrix can represent the corresponding cross-configuration vulnerability values $v_{ij}$, and the edge weights can represent the cross-segment vulnerability values $\mu_{ij}$. Of course, both perspectives are simplifications of the original risk model, and we will need to evaluate to what extent they allow us to find good solutions for the problem. Our hypothesis is that these perspectives and the use of graph coloring will allow to find good solutions in a reasonable time. In the following sections, we describe the different techniques we have evaluated to test our hypothesis. First, we discuss some reference techniques from the literature. Then, we propose two novel techniques we expect to be especially adequate for this scenario, since they take advantage of the graph structure of the problem.

### 3.3.2. Reference optimization techniques

To validate our hypothesis and evaluate the contribution of our proposed approaches, we have compared them with several optimization techniques from different scientific fields. For all these techniques, we will evaluate not only their performance in terms of optimal network configuration (i.e., risk mitigation), but also the time they take to reach a solution, which is critical for their effectiveness in real scenarios.

*Augmented Lagrangian Particle Swarm Optimization (ALPSO)*

This technique was used because particle swarm is a well-known optimizer that has been successfully applied to a number of problems [49, 50] and can be considered a generic nonlinear optimizer that uses complete information. More specifically, we have chosen a parallel augmented Lagrange multiplier particle swarm optimizer, which solves nonlinear non-smooth constrained problems using an augmented Lagrange multiplier approach to handle constraints [51].

*Augmented Lagrangian Harmony Search Optimization (ALHSO)*

We have also used a generic optimizer based on harmony search, which is an evolutionary optimization algorithm inspired by musical composition [52]. As with ALPSO, this technique will be used as a reference of a nonlinear optimizer with complete information. As above, we have used augmented lagrangian multipliers to deal with constraints.

*Hill Climber (HC)*

This technique starts from a randomly-generated graph coloring, that is, starts by assigning a random color to each component in the graph. From this point, and at every iteration, one graph node changes its color randomly. If that mutation produces a better solution, we store that coloring and go to the next iteration. This process goes on until a maximum number of iterations is reached.

*Simulated Annealing (SA)*

We have also used a widespread nonlinear optimization technique called simulated annealing (SA) [28, 53]. When a coloring yields a utility loss against the previous iteration coloring, there will be a probability for the mechanism to accept it nonetheless. This probability $P_a$ depends on the utility loss associated to the new coloring $\Delta u$, and also depends on a parameter known as annealing temperature $\tau$, so that $P_a = e^{-\Delta u/\tau}$. Annealing temperature begins at an initial value, and linearly decreases to zero throughout the successive iterations of the algorithm.

The choice of these two mechanisms (HC and SA) is not arbitrary. These types of techniques have yielded very satisfactory results in negotiation for nonlinear utility spaces [54], and are the basis for several of our previous efforts [55-62]. Furthermore, as discussed in [28], the comparison between hill-climbers and annealers allows to assess whether the scenario under consideration is a highly complex one, since in such scenarios greedy optimizers tend to get stuck in local optima, while the simulated annealing optimizer tends to escape from them.

*3.3.3. Proposed guided techniques*

The techniques described so far are generic nonlinear optimizers, which are expected to reach reasonably good results in our setting. However, they are "blind" in the sense that they do not take advantage of the network structure of the problem. In the following, we propose two novel approaches to tackle the optimization process, which take advantage of the graph structure of the problem and thus, we expect, yield better results.

*TSC-DSATUR*

This technique adapts the sequential greedy algorithm DSATUR [63]. Our implementation of TSC-DSATUR for the Threshold Spectrum Coloring problem looks for the colors minimizing the risk, as shown in the pseudocode displayed in Algorithm 1. The algorithm starts with an undefined coloring and iterates by selecting at each iteration the uncolored vertex with the highest saturation degree, that is, the one with more already-colored neighbors, the one with highest degree in the case of a tie, or a random vertex among the highest-degree vertices in the case of a double tie. Once a vertex $v$ has been selected, a color is assigned from the available set so that the interference (i.e., estimation of risk) at vertex $v$ is minimized.

```
Algorithm 1: TSC-DSATUR

Input:
    G = (V, E): network graph, with v ∈ V (set of vertices)
    C = {c_i}: available color set
    W: matrix representing interferences between colors
Output:
    S: final network graph coloring
c(v)=∅, ∀ v ∈ V
while ∃ v ∈ V / c(v)=∅ do:
    v = argmax_{x∈V; c(x)=∅} saturation_degree (x)
    c(v) = argmin_{Ci ∈ C} ∑_{u ∈ N(v); c(v)=∅} W(c(u), c_i)
end
```

For further details about this technique and its behaviour in generic graph-coloring settings, readers may refer to the work [48].

*Gossip-partitioned Belief Propagation (GPBP)*

Belief Propagation (BP) is a message-passing heuristic for solving optimization and inference problems in the context of a graphical model [64]. Under certain conditions, BP is able to find optimal solutions to factorized optimization problems, that is, optimization problems of the form

$$minimize \sum_{i \in V} \Phi_i(x_i) + \sum_{c \in C} \Psi_c(x_c)$$
$$subject\ to\ x_i \in \mathbb{R}, \forall i \in V$$

where V is a finite set of variables and C is a finite collection of subsets of V representing constraints. $\Phi_i$ functions are called variable functions (they depend on the value of a single variable), and $\Psi_i$ functions are called factor functions (they depend on specific combinations of variables called factors). All these functions are real-valued.

For the REACT purposes, we need to translate our augmented Threshold Spectrum Coloring to a factorized optimization problem. Let us see an example of how to do this for the configuration-spread projection. We use our components as variables (which can take different values $s_i$ depending on where they are redeployed) and the links between pairs of nodes (which represent similar nodes through which attack spreads may occur) as constraints. According to this, we define the corresponding functions as follows:

$$\Phi_i(s_i) = \mu_i v_i$$
$$\Psi_C(s_i, s_j) = v_{ij}\mu_{ij}, \quad \forall\ C \equiv (i, j)$$

That is, we multiply the inherent segment and configuration vulnerability values for each variable function, and we multiply the cross-segment and cross-configuration vulnerability values for each factor function. An analogous formulation could be done for the network-spread projection.

With this formulation, we try to mitigate both the impact of putting very similar components close to each other, and the impact of putting inherently vulnerable components in inherently vulnerable segments (e.g. placing a machine with a lot of accessible services in the outer DMZ). It is worth noting that this formulation differs from the TSC problem, given that here we try to minimize the sum of the contributions for all nodes in the graph, while pure TSC aims to minimize the maximum contribution for any single node in the graph. However, as we showed in [48], sum minimization is a good heuristic to minimize the maximum in this context.

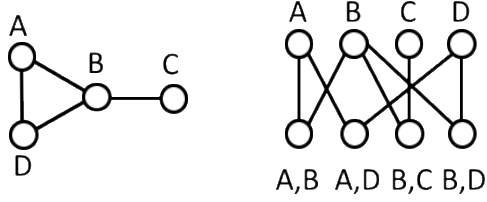

**Fig. 4.** Factor graph (right) for our example TSC problem.

Once the formulation has been established, we need to build the factor graph F of our problem, which is a bipartite graph with *variable nodes* in one side of the partition and *factor nodes* corresponding to the constraints in the other side of the partition. Links between both partitions occur between a constraint and the variable nodes it refers to. An example of the resulting factor graph F would be as shown in Fig. 4.

Finally, we would have to apply the min-sum algorithm for BP [64]. The problem with directly applying the min-sum BP algorithm to our problem is that the algorithm only has correctness and convergence guarantees when the solution is unique and the factor graph is a tree. Although solution uniqueness can be achieved with randomized weights as suggested in [64], most of our scenarios do not create tree factor graphs. Taking this into account, to ensure convergence and correctness of the algorithm, we divide the factor graph derived from our problem formulation into trees using a gossip-inspired technique [64]. Of course, when we work with the resulting set of trees, we would lose the information about the interference from neighbors in different trees. To minimize the impact of this simplification, we introduce this effect in the functions $\Phi_i(s_i)$ of the frontier nodes.

## 4. Evaluation of REACT

### 4.1. Simulation settings

For the evaluation of the proposals, we have implemented a Python simulator together with a networkX-based computational model of the problem, which realistically represents the threat model related to zero-day malware infection using diversity metrics [6] and risk-based probabilistic propagation. Unlike other papers [45], we make use of a wide range of scenarios that include very different situations, as critical infrastructures are diverse. Moreover, we have resorted to random scenarios as they are able to represent higher diversity than using a specific scenario, as we are focused on generic critical infrastructures. Finally, for the time measurements, it is worth noting that we have conducted the experiments on an Intel Core i7-2600 with 8 CPUs@3.40GHz and 8GB RAM, running Ubuntu 14.04.4 LTS.

To test the performance of the REACT system, we generated an extensive set of scenarios representing different instances of the multilayer network model described in section 3.1. In particular, we generated 60 different scenarios, as follows:

- We generated six different Asset Layer graphs, two for each of the following numbers of assets: {60, 70, 80}. Each graph was generated as an Erdős-Renyi random undirected acyclic graph, with p=0.05. This creates an average degree of dependencies between assets in the range of 3-4. That means, on average, each asset depends on 3-4 other assets. Each asset was assigned a random value, drawn from a uniform distribution between 1 and 10. Without loss of generality, dependencies between assets are assumed absolute. That is if asset $a_i$ depends on asset $a_j$ and asset $a_j$ fails, $a_i$ fails too and all its value is lost.
- For each asset graph of N assets, two component/instance layer categories have been generated, having 2N components each. Each component layer category was generated as a random undirected Erdős-Renyi graph with p values in the set {0.1,0.3}. This generates component/instance layer categories where a given zero-day is supposed to effect, on average, 10% and 30% of the components (with varying degree of similarity among instances). This models a realistic situation and a worst-case scenario. For each category, 10 different instance layers were generated corresponding to different configuration choices. For each instance layer, configurations for components were randomly drawn from a set of 10 different configurations. Inherent zero-day vulnerability values $v_i$ were drawn from a uniform distribution between 0 and 1 for each configuration. Cross-configuration vulnerability values $v_{ij}$ between configurations were set so that "further apart configurations" in the "configuration spectrum" have lower values, following an exponential decay (e.g. $v_{12} = v_{23} = \frac{1}{2}$, $v_{16} = \frac{1}{32}$, …).

- Finally, we randomly deployed each instance in a topology layer comprising 6 different network segments in a typical defense-in-depth scenario (i.e., concentric rings of security). For each pair of segments $(s_i, s_j)$, the cross-segment vulnerability value $\mu_{ij}$ depends on the distance between the segments in the topology, following an exponential decay (e.g. $\mu_{12} = \mu_{23} = \frac{1}{2}$, $\mu_{16} = \frac{1}{32}$, …). Inherent segment vulnerability values $\mu_i$ were set to $\mu_i = \frac{1}{i}$, assuming the "outer" segment is totally exposed.

For each one of the sixty scenarios, we run simulations of the REACT system in the event of the compromise of each single component, which accounts for 4200 different security breach scenarios, with each one of the four optimization techniques mentioned in section 3.3.2 and the two guided approaches proposed in section 3.3.3 to derive an alternative deployment of the components aiming to minimize risk. In all cases, we measured the reduction of the overall risk $\overline{\rho_{c_j}}$ from the original deployment to the REACT reconfiguration. We also measured the amount of time taken by the REACT system in providing a solution. We configured the REACT mechanisms to work either in the *network-spread projection* and in the *configuration-spread projection*, to see the effect of applying the same heuristics to both perspectives of the problem.

### 4.2. Simulation results for the REACT system

Figure 5 shows the risk reduction ($\overline{\rho_{c_j}}/\overline{\rho_{c_j}}^0$) for the network-spread projection and for the different techniques under study for different values of *n* and *p* with the notation (*n, p*). Note that *n* represents the number of assets (for *n* in {60, 70, 80}) and *p* is the ratio of components affected by each zero-day attack (for *p* in {0.1, 0.3}). Each bar represents the mean value and 95% confidence intervals of the risk reduction obtained from the 10 different instance layers when "re-coloring" them according to the network-spread projection (i.e., changing the configuration of the components). Moreover, risk reduction represents the ratio of the risk obtained by the optimization technique under study ($\overline{\rho_{c_j}}$) and the risk of the original deployment ($\overline{\rho_{c_j}}^0$), and, for that reason, lower values represent better performance. Additionally, for each scenario category, we show as a solid black line the mean value for all the techniques under study. Results show that, in terms of risk reduction the best performance is obtained by GPBP, closely followed by TSC-DSATUR. Our proposals highly outperform the other optimizers. This is probably due to the fact that, being generic optimizers, they do not take advantage of the network structure of the problem. On the other hand, in Fig. 6 we show a similar plot but for the configuration-spread projection, that is, redeploying components in different network segments without altering their configuration. For this projection, the performance of the different algorithms under study is closer than in the network-spread projection. More specifically, the performance of our proposal GPBP is the best, but now TSC-DSATUR takes second place in three of the categories, and SA takes second place in the other three.

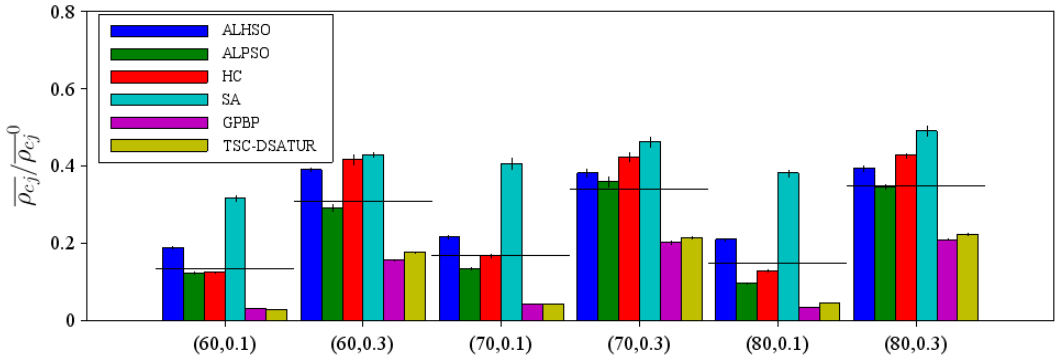

**Fig. 5.** Risk reduction results for the network-spread projection.

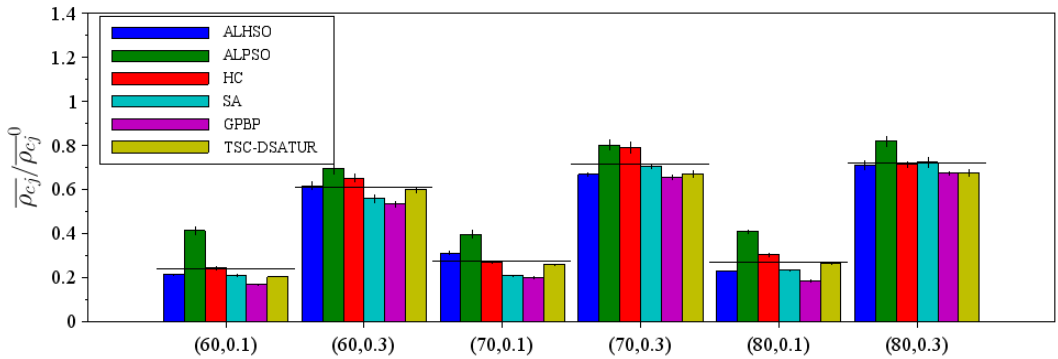

**Fig. 6.** Risk reduction results for the configuration-spread projection

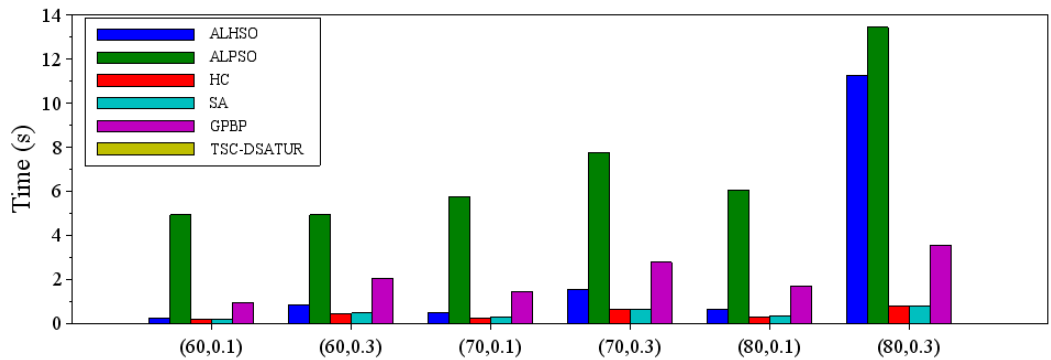

**Fig. 7.** Time required by REACT to recover from an incident using different techniques for the network-spread projection.

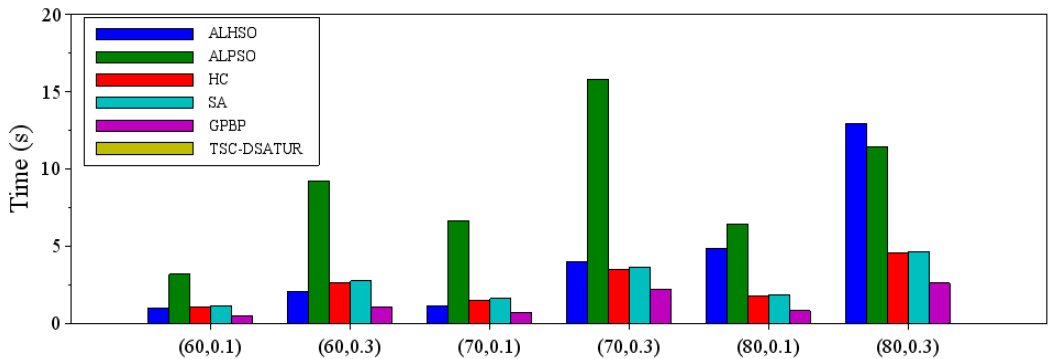

**Fig. 8.** Time required by REACT to recover from an incident using different techniques for the configuration-spread projection.

It is also very interesting to compare the results reached by using both projections: network and configuration-spread (Fig. 5 and 6, respectively). If we compare the reduced risk for a certain category ($n, p$) and for a certain technique in both figures, we can clearly conclude that using the network-spread projection we are able to reduce the risk much better than using the configuration-spread projection.

Since REACT is a framework that must operate in a timely manner to minimize the effect of possible intrusions, we have also to evaluate the time needed by each of the algorithms to be reduce the risk. For that purpose, Figs. 7 and 8 represent the running times for each technique and for the network-spread and configuration-spread projections, respectively. Note that the time needed to execute TSC-DSATUR is negligible, so it cannot be distinguished in the figures. The second fastest technique depends on the projection under use, because in the configuration-spread projection it is GPBP while in the network-spread projection there are two very similar techniques (HC and SA), which yield significantly better times than GPBP. Taking into account the results about the performance of the different techniques in terms of risk reduction and running time, we can conclude that, from all the techniques under study, the best choice will depend on the time constraints we have. If we want to obtain the best performance in terms of resiliency, we should definitely use GPBP. However, if time limits are very tight, probably the best choice is to use TSC-DSATUR, as its performance in terms of risk reduction is close to the best one (GPBP), but it is the fastest, by far, of all the approaches we studied.

Now, we focus on the effect that some network metrics have in the achievable risk reduction. In Fig. 9 describes how the risk reduction changes as a function of the expected average degree of the network (product $np$). In this figure we show that the risk reduction depends on $np$ almost linearly for all the techniques under study. From this figure we can conclude that the reachable risk reduction is higher for simpler graphs than for the most highly connected. The results shown are for the configuration-spread projection, although similar ones are obtained for the alternative network-spread projection.

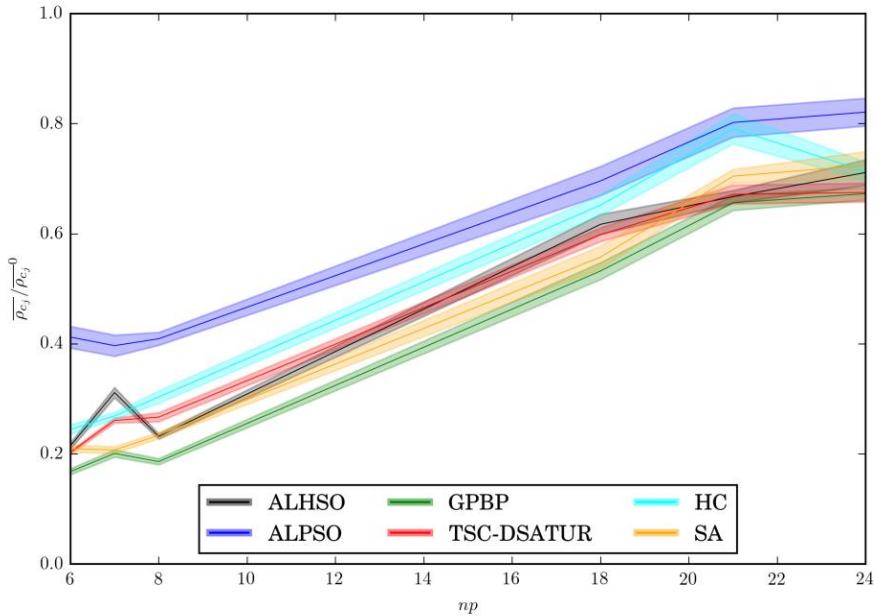

**Fig. 9.** Risk reduction as a function of the product $np$.

Finally, we are interested in how different network average centrality metrics affect the risk reduction for the different techniques. This effect is shown in Fig. 10 (again, for the configuration-spread projection, although similar conclusions can be drawn from the network-spread projection). In general, centrality measures the importance of a vertex (node) in a graph [65]. This importance can be measured in different terms, which leads to different centrality metrics. *Betweenness* measures to which extent a vertex in the network is on the shortest paths that join other vertices. More specifically, for each pair of vertices $s$, $t$ different from $v$, the ratio of shortest paths between $s$ and $t$ containing $v$ is obtained, and these ratios are summed up. *Closeness* centrality represents the mean distance from a node to other nodes, so it can be computed as the inverse of the farness normalized by the number of other nodes. *Degree* centrality accounts for the number of neighbors a vertex has. Finally, eigenvector centrality can be considered an extension of the degree centrality, but in eigenvector centrality a node is important depending not in the number of neighbors, but in the importance of its neighbors. Storing the centralities of the vertices in a vector, this turns out to be the eigenvector associated to the largest eigenvalue of the adjacency matrix of the graph. As it can be deduced from Fig. 10, the behavior among the different techniques do not differ depending on the metrics, so we can perform an analysis for all the techniques as a whole. It is interesting to see that the best risk reductions can be reached for the lowest values of closeness and degree centrality. However, it is desirable to have high values of betweenness centrality for REACT to be able to reduce risk more drastically. Finally, and regarding the eigenvector centrality, we can conclude that this metric does not influence in the risk reduction, as we have very different reductions for the same values of this centrality. To summarize the results in terms of the underlying physical infrastructure on the effectiveness of the system, we can conclude that for having the best resilience it is of paramount importance to have low values for the network degree (from Fig. 9), closeness and degree centralities (from Fig. 10) and high values for the betweenness centrality (from Fig. 10).

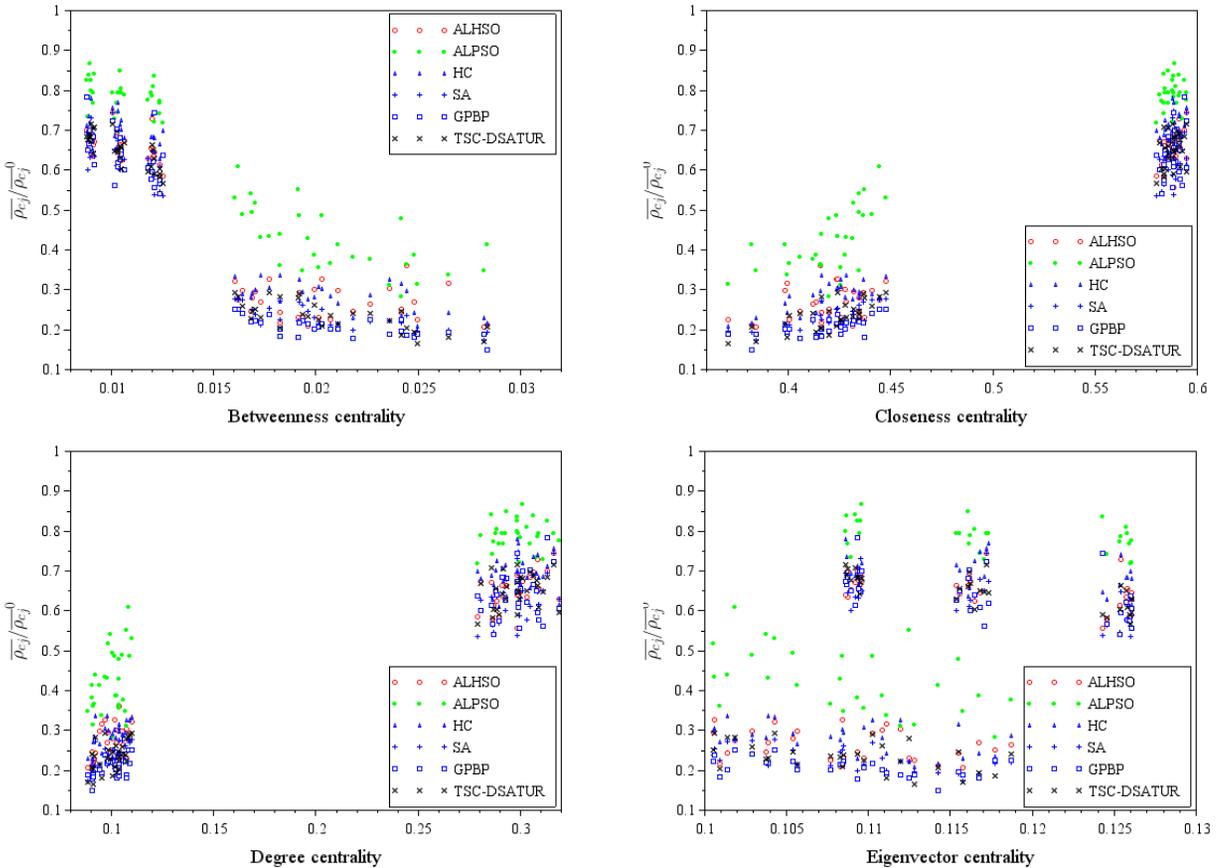

**Fig. 10.** Effect of graph metrics in risk reduction.

*4.3. Comparison of TSC-DSATUR with other relevant approaches in the literature*

As stated in Section 2.3, our work [44] is pioneer in proposing a reactive redeployment of components to reduce the risk after an attack has occurred. In this paper, we have generalized the approach in [44] with both a more detailed model and the use of the graph coloring technique TSC-DSATUR, which we had tested in graph coloring problems in a different domain [48]. As described in Section 2.3, in a very recent work [45], other authors proposed the use of graph coloring techniques in a related setting, to increase the security of a Smart Grid installation by adding diversity in the deployment of security mechanisms. There are some key differences between their work and ours:

- Their model is specific to Smart Grid settings, while ours is generic for any critical network infrastructure.
- Their model only considers the optimization of the configuration of security mechanisms, while in our model we propose to optimize the configuration of all components of the infrastructure.
- Their model is intended for design instead of reaction, although this per se does not limit its applicability for reactive resilience.
- Their model does not allow to optimize the location where the different components are deployed. This makes it only comparable to our network-spread projection.
- Their proposed approach, which they call Graph Coloring Game (GCG) uses only proper colorings, i.e., colorings with no monochromatic edges, and therefore they assume the number of available configurations is enough for such colorings to exist, making the number of configurations available depend on the structure of the scenario graph under consideration (in particular, of the maximum degree $\Delta(G)$ of the graph G). In our model, we assume the number of available colors (which represent configurations or network segments, depending on which projection is considered) to be limited, and we allow for the use of improper colorings (i.e., colorings where there are monochromatic edges).

Taking these key differences into account, and in the spirit of making a fair comparison in spite of them, we have performed a second set of experiments, where we have compared GCG with TSC-DSATUR, which, although not being the best of our proposed approaches, is the most directly comparable one, since they are both graph coloring techniques. We have also included two classic techniques for proper colorings as references: the Welsh-Powell coloring (WP) [66] and the DSATUR coloring [63]. We have assumed GCG to be capable of optimizing the configuration of all components in the network, and not only security mechanisms. Finally, we have run the three proper coloring techniques (GCG, WP, and DSATUR) in all of our scenarios with an unlimited set of colors (so that they could achieve proper colorings), and then we have run TSC-DSATUR with the minimum number of colors used by these techniques at each scenario. This is the worst possible case for TSC-DSATUR.

Table 1 summarizes the results of the comparison of the abovementioned techniques (DSATUR, WP and CGC) with our proposal (TSC-DSATUR). In this table we show the risk reduction $\overline{\rho_{c_J}}/\overline{\rho_{c_J}}^0$ (average and 95% confidence intervals) that the different proposals are able to reach, together with the computation time required by each one. Note that we show in bold the best value (i.e., the lowest) obtained for each scenario. Analyzing the results, we show that TSC-DSATUR obtains the best results in terms of risk reduction, followed by GCG. Note that for *(n,p)=(60,0.3)* CGC is slightly better than TSC-DSATUR, although the difference is not statistically significant. Results also show that WP is usually better than DSATUR, which usually gets the worst performance among all the techniques under study. It is also interesting to compare the computation time required by the different techniques. Although TSC-DSATUR is the slowest technique among all the studied ones, it must be noted that its computation time is 5ms in the largest scenario under consideration. Finally, Fig. 11 shows the risk reduction obtained by the different techniques as a function of the product np. Note that the shaded areas around the different curves represent the 95% confidence intervals. Results show again that the best choice for reducing risk, from all the studied techniques, is TSC-DSATUR.

| $n$ | $p$ | Approach | $(\overline{\rho_{c_j}}/\overline{\rho_{c_j}}^0)$ Average | CI | Time |
|---|---|---|---|---|---|
| 60 | 0.1 | DSATUR | 0.6798 | 0.0192 | 1ms |
| | | WP | 0.6649 | 0.0174 | <1ms |
| | | GCG | 0.5810 | 0.0151 | <1ms |
| | | TSC-DSATUR | **0.3549** | 0.0113 | 1ms |
| | 0.3 | DSATUR | 0.7674 | 0.0073 | 1ms |
| | | WP | 0.6978 | 0.0057 | <1ms |
| | | GCG | **0.6263** | 0.0125 | <1ms |
| | | TSC-DSATUR | 0.6406 | 0.0181 | 3ms |
| 70 | 0.1 | DSATUR | 0.8199 | 0.0149 | 1ms |
| | | WP | 0.7546 | 0.0144 | <1ms |
| | | GCG | 0.7026 | 0.0160 | <1ms |
| | | TSC-DSATUR | **0.4741** | 0.0069 | 2ms |
| | 0.3 | DSATUR | 0.8704 | 0.0136 | 2ms |
| | | WP | 0.8331 | 0.0120 | <1ms |
| | | GCG | 0.7490 | 0.0086 | <1ms |
| | | TSC-DSATUR | **0.5298** | 0.0146 | 4ms |
| 80 | 0.1 | DSATUR | 0.7796 | 0.0169 | 2ms |
| | | WP | 0.8467 | 0.0192 | <1ms |
| | | GCG | 0.6218 | 0.0076 | <1ms |
| | | TSC-DSATUR | **0.4733** | 0.0081 | 2ms |
| | 0.3 | DSATUR | 0.8153 | 0.0121 | 2ms |
| | | WP | 0.7867 | 0.0131 | <1ms |
| | | GCG | 0.7560 | 0.0161 | <1ms |
| | | TSC-DSATUR | **0.6127** | 0.0053 | 5ms |

Table 1. Comparison with previous proposals.

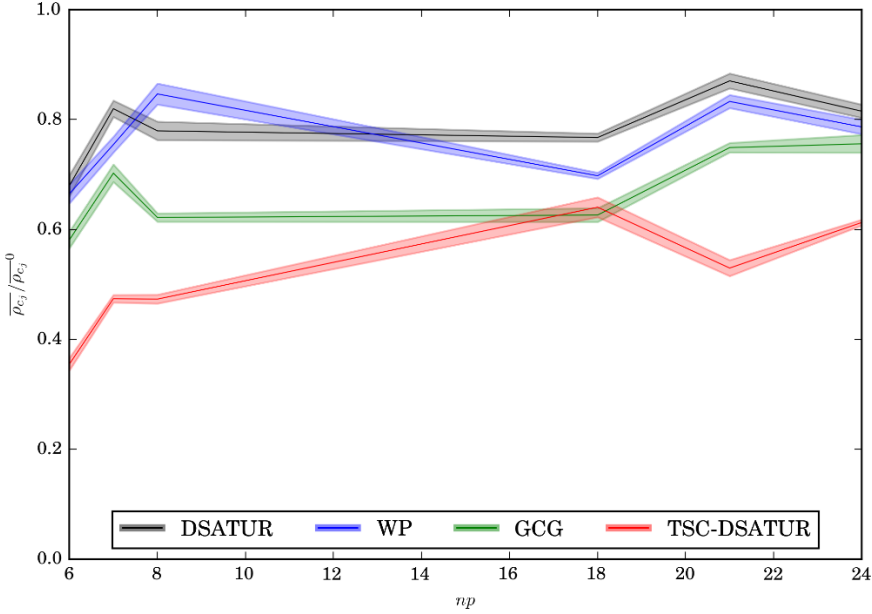

Fig. 11. Comparison of the risk reduction with previous proposals as a function of the product $np$.

## 5. Conclusions and future work

Critical network infrastructures (CNIs) are key components in current society so we must protect them from malicious attacks that can exploit their vulnerabilities and cause cascading failures with catastrophic consequences. In this work we propose REACT, a framework for reactive network resilience in (semi-)virtualized CNIs over a zero-day attack threat model. REACT provides a network with the ability to quickly reconfigure itself in the event of a security incident so that the risk of further damage is mitigated.

The main contributions of the paper are the following. First, we propose a model for risk reduction under zero-day attacks based on multi-layer graphs. Second, since resilience must be achieved in a timely manner, we propose two different perspectives to the problem, and use them to approximate it as an extension of the well-known graph coloring problem. Third, we propose two novel techniques, GPBP and TSC-DSATUR, to solve the problem and compare them with several state-of-the-art optimization techniques: a particle swarm optimizer, a harmony search optimizer, a hill climber and a simulated annealer. Results show that our proposals outperform the other techniques in terms of risk reduction and execution time, which key for fast resilience that limits the damages after a malicious attack. In particular, GPBP obtained the best results in terms of risk reduction while TSC-DSATUR, was able to reduce risk almost as much as GPBP, but much more quickly. Regarding the network metrics, we can conclude that some of them clearly affect the achievable risk reduction, so it is important that they are taken into account in network design. Finally, we show some of the features that the underlying multilayer network layout must have to be able to highly reduce risk.

Although our proposed techniques have achieved very satisfying results, in a future we are interested in dealing with network resilience under multiple simultaneous malicious attacks, as this framework will be able to represent more accurately some realistic scenarios. We are also working to integrate both projections (*network-spread* and *configuration-spread*) in the optimization process, so that we can achieve higher risk reductions. Finally, we want to study the influence of multilayer network metrics in a similar way as we have done in this work with graph metrics over the projection graph.

## Acknowledgements

All the authors are supported by MINECO Projects TIN2014-61627-EXP and TIN2016-80622-P (AEI/FEDER, UE), and by the University of Alcalá project CCG2016/EXP-048. In addition, David Orden is supported by Project MTM2017-83750-P of the Spanish Ministry of Science (AEI/FEDER, UE), as well as by the European Union's Horizon 2020 research and innovation programme under the Marie Sklodowska-Curie grant agreement No 734922.